\documentstyle[aps,prb,tighten,12pt]{revtex}

\begin{document}

\draft

\title{Analytical solution of generalized Burton--Cabrera--Frank
equations for growth and post--growth equilibration on vicinal
surfaces}
\author{S. Harris\cite{byline1}} \address{Department of Civil
Engineering and Operations Research,\\ Princeton University,
Princeton, New Jersey 08544}
\author{Pavel \v{S}milauer\cite{byline2}}
\address{Interdisciplinary Research Centre for Semiconductor
Materials, The Blackett Laboratory, Imperial College, London
SW7~2BZ, United Kingdom}

\date{\today}

\maketitle

\begin{abstract}

We investigate growth on vicinal surfaces by molecular beam epitaxy
making use of a generalized Burton--Cabrera--Frank model. Our
primary aim is to propose and implement a novel analytical program
based on a perturbative solution of the non--linear equations
describing the coupled adatom and dimer kinetics. These equations
are considered as originating from a fully microscopic description
that allows the step boundary conditions to be directly formulated
in terms of the sticking coefficients at each step. As an example,
we study the importance of diffusion barriers for adatoms hopping
down descending steps (Schwoebel effect) during growth and
post-growth equilibration of the surface.

\end{abstract}


\section{Introduction}
\label{sec:intro}

Growth on vicinal surfaces, with a misorientation angle of  few
degrees, is of profound importance as a method of preparing
high--quality epitaxial layers and heterostructures.\cite{tt} This
is also frequently used to obtain insight into the fundamental
kinetics of molecular beam epitaxy (MBE) growth.\cite{n,s1} The
renewed interest in the analytical model of Burton, Cabrera, and
Frank\cite{BCF} (BCF) describing epitaxial growth on vicinal
surfaces is thus not surprising. In the BCF model, a crystal grows
by step propagation due to incorporation of atoms deposited onto the
terraces. No interactions between the atoms are considered, and it
is also supposed that adatom concentrations at the step edges are at
equilibrium and the effect of the step movement can be neglected.
This model has been recently generalized for applications pertaining
to growth by MBE.\cite{step,gi,mb1,mb2,fuen} The
far--from--equilibrium character of MBE has led to lifting all of
the above restrictions and taking into account the influence of the
step movement,\cite{step} deviations from local equilibrium at the
step edges,\cite{gi} and, finally and most importantly, effects
related to lateral interactions of adatoms on the terraces, in
particular formation of adatom islands.\cite{mb1,mb2,fuen}

Our primary concern in this paper is to propose and carry out an
analytical program for solving the generalized BCF equations
including non--linear terms due to adatom interactions. The
non--linear structure of these equations has been the major obstacle
to obtaining analytical solutions and for this reason only numerical
solutions\cite{mb1,mb2,fuen} have been reported so far. The steady
state solution of these equations\cite{sh1} was previously
considered through the use of a microscopic formulation based on
adatom and dimer position--velocity distribution functions,
$f(x,v)$, ${\cal F}(x,v)$, instead of the macroscopic densities
$m(x)$,
$M(x)$ more commonly used,\cite{BCF,gi,mb1,mb2,fuen} and we will
implicitly use this formulation here as well. However, as discussed
below, here this distinction is only operationally significant in
the treatment of the boundary conditions which we are able to
describe in terms of the step sticking coefficients.

We will assume that the growth conditions (substrate temperature,
growth rate, etc.) are such that step flow dominates so that only
small islands (dimers) have to be considered, but the step movement
is still sufficiently slow so that the steps can be considered
stationary in solving the boundary value problem.\cite{mb1,mb2}
These restrictions are imposed only as simplifications and their
removal would increase the complexity and decrease the clarity of
our results. All other assumptions
introduced in this paper will be clearly noted and assessed in the
following sections. While it is our belief that some of these could
be eliminated, our main goal here is to obtain a closed form
analytical solution that is based on sound physical and mathematical
reasoning.

The outline of this paper is as follows. The necessary equations and
notation are presented in Sec.~\ref{subsec:basic_eq} where the
distinction between the microscopic and macroscopic formulations is
made explicit. Our procedure for the solution of these equations
then follows in Sec.~\ref{subsec:plan} and this is implemented in
Sec.~\ref{sec:sol}. As a demonstration of our results, we study the
effects of barriers to hopping down descending step edges (Schwoebel
effect) in Sec.~\ref{sec:recovery}. Sec.~\ref{sec:summary} is a
brief summary of the results of the paper.

\section{Theoretical Considerations}

\subsection{Basic Equations}
\label{subsec:basic_eq}

We consider a stepped crystal substrate on which adatoms are
uniformly deposited at a rate $Fa^2$ where $F$ is the flux
(m$^{-2}$s$^{-1}$) and $a$ the lattice constant (see
Fig.~\ref{fig:model}). As indicated in the introductory section, the
temperature and the flux are supposed to be such that only dimer
formation has to be considered but the steps are still sufficiently
slow and can be taken as stationary in solving the boundary value
problem.\cite{step-flow} Equations describing the evolution of the
adatom and dimer densities have been presented and used by several
authors.\cite{mb1,mb2,fuen} These same equations can also be
obtained from the microscopic description used by us earlier in
treating the steady state problem\cite{sh1} if we also assume that
the dimers are immobile and that the adatom current is given by
Fick's law. The last assumption, which is an integral part of the
macroscopic description, is equivalent to neglecting an initial
stage in which the current changes over the time scale
$\tau_{0}$ (see below) from zero at the time $t\!=\!0$ to the Fick's
law value. This should not have a significant qualitative or
quantitative impact on our results and therefore seems justified at
this juncture of the theoretical development.

Denoting the adatom and dimer densities by $m(x,t)$, $M(x,t)$ to
indicate that they are the result from averaging the distribution
functions $f$,
$\cal F$ over velocity, we then have\cite{mb1,mb2,sh1}

\begin{eqnarray}
\frac{\partial m}{\partial t} &=& D\frac{\partial^{2} m}{\partial
x^{2}} + F + 2\gamma M - (8/\tau)m - 2\Gamma m^{2} \label{eq:m_t} \\
\frac{\partial M}{\partial t} &=& (4/\tau)m + \Gamma m^{2} - \gamma M
\label{eq:M_t} \\  0&\le&x\le L~,~~~t>0~~~,\nonumber
\end{eqnarray}

\noindent where $D$ is the diffusion coefficient, $\tau\!=\!1/Fa^2$,
$\gamma\!=\!\tau_{0}^{-1} \exp(-E_{N}/k_{B}T)$ with the
characteristic time for adatom surface diffusion
$1/\tau_{0}\!=\!2D/a^2\!=\!\nu_0\,\exp(-E_{D}/k_{B}T)$,
$\nu_0$ being the attempt frequency, $E_{N}$ the dimer binding
energy and
$E_{D}$ the surface diffusion barrier for a free adatom, and $T$ is
the temperature and $k_{B}$ is the Boltzmann's constant. Also, we
have taken the number of dimer--forming sites around a free adatom as
four.\cite{fuen,ori}

The terms on the right--hand side of Eq.~(\ref{eq:m_t}) represent
(from left to right): diffusion of free adatoms, deposition of an
atom on a surface site with all nearest--neighbor positions
unoccupied, dimer break--up, dimer formation due to direct collision
of an atom from the beam with a surface adatom, and dimer formation
when two diffusing adatoms meet each other (cf.
Fig.~\ref{fig:model}). The prefactor $\Gamma\!=\!\sigma
a^{2}/\tau_{r}$ with
$\sigma$ the capture efficiency which we subsequently set to
unity,\cite{mb1} combines a geometrical factor $\sigma a^{2}$ and a
relaxation time that we have denoted as $\tau_{r}$. In previous
numerical studies\cite{mb1,mb2,fuen} $\tau_{r}\!=\!\tau_{0}$ has
been used; here we will use $\tau_{r}\!=\!(FD)^{-1/2}$ which is a
considerably larger quantity. It can be shown\cite{sh2} that
$\tau_{r}\!=\!(FD)^{-1/2}$ is in quantitative agreement with
Eqs.~(\ref{eq:m_t}), (\ref{eq:M_t}) while
$\tau_{r}\!=\!\tau_{0}$ is not. Separate justifications based on
kinetic arguments have also been given in the literature.\cite{till}

The initial conditions for Eqs.~(\ref{eq:m_t}), (\ref{eq:M_t}) are
$m(x,0)\!=\!M(x,0)\!=\!0$, {\it i.e.\/} the beam is initiated at
$t\!=\!0$ at which time the terraces are unoccupied. So far there is
no difference between our description, Eqs.~(\ref{eq:m_t}),
(\ref{eq:M_t}) with the above initial conditions, and the
macroscopic descriptions cited earlier.\cite{mb1,mb2,fuen} However,
in specifying the boundary conditions the advantages of the
microscopic theory become clear and a major difference between the
two approaches emerges. At the level of approximation in the
microscopic theory\cite{sh1,sh3} leading to the above kinetic
equations, the distribution function $f$ describing the adatoms is
given as

\begin{equation}
\label{eq:distrib} f(x,v,t) = \biggl\{m_1(x,t)\Theta(v) +
m_2(x,t)\bigl[1-\Theta(v)\bigr]
\biggr\} \, (2\pi\alpha)^{-1/2}\,{\rm e}^{-v^2/2\alpha}
\end{equation}

\noindent where $\Theta(v)$ is the standard Heaviside function and
$\alpha\!=\!k_{B}T$ (the adatom mass is expressed in units in which
it is unity). This leads to

\begin{equation}
\label{eq:m&j} m = (m_1+m_2)/2~,~~j = -D\frac{\partial m}{\partial
x} = (\alpha/2\pi)^{1/2}(m_1-m_2)
\end{equation}

\noindent It follows from Eqs.~(\ref{eq:distrib}) and (\ref{eq:m&j})
that
$m_1$ is the density of adatoms moving with positive velocity on the
terrace and $m_2$ the density of those moving with negative
velocity. This level of detail is particular to the microscopic
description and allows us to prescribe natural boundary conditions.
The most general of these, which we use here, is given in terms of
the step sticking coefficient $S_i$,
$i\!=\!0,L$ for $x\!=\!0,L$, respectively (Fig.~\ref{fig:model}). For
asymmetric step conditions at $x\!=\!0,L$

\begin{equation}
\label{eq:s_coeff} S_0 = \left[m_2(0)-m_1(0)\right]/m_2(0)~,~~S_L =
\left[m_1(L)-m_2(L)\right]/m_1(L)~~,
\end{equation}

\noindent with $0\!\le\!S_i\!\le\!1$. The finite parameters $S_i$
describe the full range of step conditions from blocking
($S_i\!=\!0$) to total absorption ($S_i\!=\!1$). In addition to the
conceptual advantage in using the microscopic boundary condition
there is also an operational advantage. From Eq.~(\ref{eq:m&j}) we
find

\begin{mathletters}
\label{eq:bound}
\begin{eqnarray} -h_Lm & = & \frac{\partial m}{\partial x}~,~~x\!=\!L
\label{eq:bound_a}\\  h_0m & = & \frac{\partial m}{\partial
x}~,~~x\!=\!0
\label{eq:bound_b}
\end{eqnarray}
\end{mathletters}

\noindent with
$h_i\!=\!\left[S_i/D(2-S_i)\right](2\alpha/\pi)^{1/2}$. Thus, for the
full range of step kinetics the boundary condition is of the Neumann
(or radiation) type and a single solution depending on the $S_i$
describes all possible sticking conditions. Eqs.~(\ref{eq:m_t}),
(\ref{eq:M_t}), and (\ref{eq:bound}) together with the initial
conditions define the problem to be considered in mathematical
terms. Note that there is no boundary condition for $M$ required for
the case considered of slow steps and fixed dimers. In the steady
state an exact solution for these equations was obtained\cite{sh1}
despite their non--linearity; here we cannot expect to be so
fortunate and therefore we have to look for approximate means of
solution.

\subsection{Plan of Solution}
\label{subsec:plan}

We propose to solve the problem defined above by tailoring a
procedure of solution to our intuitive expectation of how the
physical process evolves. Initially the terraces are unoccupied,
adatoms are deposited and begin to diffuse to the steps and, as
their density increases, dimers begin to be formed and, finally, the
latter may begin to disassociate. Thus, three separate evolutionary
stages, most likely not sharply defined, are suggested. One
mathematical interpretation of this is to consider the terms in
Eq.~(\ref{eq:m_t}) representing dimer effects to be of higher order
in an approximation scheme; $M$ also will be of higher order. Writing
$m\!=\!m^{(0)}\!+\!m^{(1)}\!+\ldots$~,~~$M\!=\!M^{(1)}\!+\!M^{(2)}\!+\ldots$
and re-writing Eq.~(\ref{eq:m_t}) we have

\begin{eqnarray}
\frac{\partial m}{\partial t} & =  & D \frac{\partial^2 m} {\partial
x^2} + F - R(m,M) \eqnum{$1^\prime$} \\
\frac{\partial M}{\partial t} & =  & \frac{1}{2} R(m,M)
\eqnum{$2^\prime$}
\end{eqnarray}

\noindent ($R(m,M)$ represents all the terms on the right--hand side
of Eq.~(\ref{eq:m_t}) which are not given explicitly) so that our
procedure for solution is implemented by considering the following
equations,

\begin{eqnarray}
\frac{\partial m^{(0)}}{\partial t} & =  & D\frac{\partial^{2}
m^{(0)}}{\partial x^{2}} + F \label{eq:m0_t} \\
\frac{\partial m^{(1)}}{\partial t} & =  & D\frac{\partial^{2}
m^{(1)}}{\partial x^{2}} - R(m^{(0)},0) \label{eq:m1_t} \\
\frac{\partial M^{(1)}}{\partial t} & =  & \frac{1}{2}
R(m^{(0)},M^{(1)})~,~~\ldots \label{eq:M1_t}
\end{eqnarray}

\noindent with, e.g.

\begin{equation} h_0m^{(0)} = \frac{\partial m^{(0)}}{\partial
x}~,~~h_0m^{(1)} =
\frac{\partial m^{(1)}}{\partial x}~,~~\ldots~~~,~~x\!=\!0
\label{eq:bound_2} \end{equation}

\noindent The connection between the preceding physical and
mathematical descriptions becomes clearer if we operate on
Eq.~(\ref{eq:m_t}) with
$\partial /{\partial t}$ and use Eq.~(\ref{eq:M_t}) to obtain

\begin{eqnarray}
\frac{\partial}{\partial t}\left[ \frac{\partial m}{\partial t} - D
\frac{\partial^{2} m}{\partial x^{2}} - F \right] & =  & -\gamma
\left[
\frac{\partial m}{\partial t} - D \frac{\partial^{2} m}{\partial
x^{2}} - F \right] - \frac{\partial}{\partial t}\left[ 8m/\tau +
2\Gamma m^2 \right]
\label{eq:full} \\  & \equiv & -R_1(m) - \frac{\partial
R_2(m)}{\partial t} \label{eq:1dprime}
\eqnum{$1^{\prime\prime}$} \end{eqnarray}

\noindent The first term on the right--hand side is the contribution
due to dimer break--up, indicated by the $\gamma$ prefactor, while
the second is the contribution due to dimer formation. Since
$R_1(m^{(0)})\!=\!0$ by virtue of Eq.~(\ref{eq:m0_t}) and ${\partial
R_2(m^{(0)})}/\partial t\!\neq\!0$ we see that the mathematical
description given by Eqs.~(\ref{eq:m0_t}), (\ref{eq:m1_t}) reflects
our conjectured physical picture in which dimer break--up effects
are of higher order than dimer formation effects. In lowest order
the right--hand side of Eq.~(\ref{eq:full}), which contains only
dimer effects, is set equal to zero and the left--hand side used to
determine $m^{(0)}$, which identically recovers Eq.~(\ref{eq:m0_t})
since ${\partial F}/\partial t\!=\!0$, $\partial
m^{(0)}(x,0)/\partial t\!=\!F$,
$\partial^2m^{(0)}(x,0)/{\partial x^2}\!=\!0$ so that

\begin{mathletters}
\label{eq:new}
\begin{eqnarray}
\frac{\partial}{\partial t}\left[\frac{\partial m^{(0)}}{\partial t}
- D
\frac{\partial^2 m^{(0)}}{\partial x^2} - F\right] & = &
\frac{\partial}{\partial t}\left[\frac{\partial m^{(0)}}{\partial t}
- D
\frac{\partial^2 m^{(0)}}{\partial x^2}\right]~=~0~,~~{\rm and}
\label{eq:new_a}\\
\frac{\partial m^{(0)}(x,t)}{\partial t}& = & D\frac{\partial^2
m^{(0)}(x,t)}{\partial x^2}~+~F  \label{eq:new_b}
\end{eqnarray}
\end{mathletters}

\noindent In what follows we will directly use Eqs.~(\ref{eq:m0_t})
and (\ref{eq:m1_t}) to determine $m^{(0)}$, $m^{(1)}$.

The preceding discussion contains a rationale, based on physical
reasoning, together with a supporting  mathematical argument, for
implementing a perturbative approximation of Eqs.~($1^\prime$),
($2^\prime$). Our reasoning here is admittedly heuristic, however
given the intractable nature of reaction--diffusion equations in
general (a bounded domain is the most difficult case), we believe
that the approach outlined above is a reasonable first step in
obtaining a useful analytical solution for the generalized BCF
equations in the regime near step flow where dimer effects begin to
play a role. It has been shown\cite{mb2} that in this regime the
moving boundary effect can still be neglected so that this omission
here is not an issue. It is possible that the procedure outlined
above could be formalized through the use of a scaling argument
leading to the introduction of one or more parameters of smallness;
we have not attempted to do this here since the underlying physical
rationale is so intuitively appealing and transparent. In the next
section we obtain explicit results for $m^{(0)}$, $m^{(1)}$, and
$M^{(1)}$ by implementing the program described above.

\section{Solutions}
\label{sec:sol}

\subsection{Lowest Order; $m^{(0)}$}

As discussed above, we write $m\!=\!m^{(0)}\!+\!m^{(1)}\!+\ldots$ and
determine $m^{(0)}$, $m^{(1)}$ from Eqs.~(\ref{eq:m0_t}),
(\ref{eq:m1_t}) together with the boundary conditions specified by
Eq.~(\ref{eq:bound_2}). These equations can be solved by standard
methods;\cite{cj} we first consider the equation for $m^{(0)}$,
Eq.~(\ref{eq:m0_t}) for the general case where the beam $F$ is
turned off at $t\!=\!t^\star$. The exact solution for
$t\!<\!t^\star$ is

\begin{equation}
\label{eq:m0_sol} m^{(0)}(x,t) = \sum_{n=1}^{\infty} (F/Db_n^2)
\left(1-{\rm e}^{-b_n^2Dt}\right) K(b_n,x) \int_0^L K(b_n,x^\prime)
{\rm d}x^\prime
\end{equation}

\noindent where the eigenfunctions $K(b_n,x)$ are\cite{sh3}

\begin{equation}
\label{eq:eigen_K} K(b_n,x) = \sqrt{2}\left[b_n \cos b_nx + h_0 \sin
b_nx\right]
\left[(b_n^2+h_0^2)(L+h_L(b_n^2+h_L^2)^{-1}+h_0)\right]^{-1/2}
\end{equation}

\noindent and the eigenvalues $b_n$ are given by the positive roots
of
$b(h_0+h_L)\!=\!(b^2-h_0h_L)\tan bL$. The terms without the
exponential factor can be directly summed but the tedious algebraic
manipulation required can be avoided by noting that these terms are
the non--decaying part of the solution, {\it i.e.\/} the steady
state solution of Eq.~(\ref{eq:m0_t}). In order to simplify the
remaining sum and obtain a result that is both easier to use and
also makes the next level of approximation tractable we will replace
$b_n$, {\it in the exponential term only\/}, by a lumped eigenvalue,
$b$.\cite{boltz} In Sec.~\ref{subsec:eigenval} we indicate a
procedure for determining $b$ in terms of the system parameters
through the imposition of a consistency constraint.

Making the above simplification we then have

\begin{equation}
\label{eq:m0_simpl} m^{(0)}(x,t) = (F/2DA_0)
\left(-A_0x^2+h_0B_0x+B_0\right) {\cal F}_0(t)
\end{equation}

\noindent where

\begin{mathletters}
\label{eq:coeff_m0}
\begin{eqnarray} A_0&=&h_0+h_L+h_0h_LL \label{eq:coeff_m0_a} \\
B_0&=&2h_0L+h_0h_LL^2 \label{eq:coeff_m0_b} \\ {\cal
F}_0(t)&=&1-{\rm e}^{-b^2Dt}~,~~t<t^\star \label{eq:coeff_m0_c} \\
&=&{\rm e}^{-b^2D(t-t^\star)}-{\rm e}^{-b^2Dt}~,~~t>t^\star
\label{eq:coeff_m0_d}
\end{eqnarray}
\end{mathletters}

\noindent The above results can be non--dimensionalized (see
Appendix~\ref{app:nondim}) but for now it is most convenient to
retain the dimensional form shown above.

\subsection{First Order: Adatom Density $m^{(1)}$}

The equation for $m^{(1)}$ is given from Eq.~(\ref{eq:m1_t}) as

\begin{equation}
\label{eq:m1_t_2}
\frac{\partial m^{(1)}}{\partial t} - D \frac{\partial^2
m^{(1)}}{\partial x^2} = -(8/\tau)m^{(0)} - 2\Gamma(m^{(0)})^2 =
R_2(m^{(0)})
\end{equation}

\noindent This equation is a more general version of
Eq.~(\ref{eq:m0_t}), {\it i.e.\/} a diffusion equation with a source
term that here is not constant but depends on both space and time.
The formal solution is

\begin{equation}
\label{eq:m1_sol} m^{(1)} = \sum_{n\!=\!1}^{\infty} {\rm
e}^{-b_n^2Dt} K(b_n,x) \int_0^t {\rm d}t^\prime {\rm
e}^{b_n^2Dt^\prime} \int_0^L{\rm d}x^\prime K(b_n,x^\prime)
R_2(m^{(0)}(x^\prime,t^\prime))
\end{equation}

\noindent We denote the contribution from that part of $R_2$ linear
in
$m^{(0)}$ as $m^{(11)}$ so that $m^{(1)}\!=\!m^{(11)}\!+\!m^{(12)}$;
further we note that $m^{(11)}$ will be identically zero when the
beam is turned off. Using Eq.~(\ref{eq:m0_simpl}) for $m^{(0)}$ and
again replacing
$b_n$ by $b$ in the exponential terms after the $t^\prime$
integration is carried out, we find

\begin{equation}
\label{eq:m11} m^{(11)}(x,t) = \frac{4F}{\tau D^2A_{11}}
\left( \frac{-A_{11}}{12}x^4 + \frac{h_0B_{11}}{6}x^3 +
\frac{B_{11}}{2}x^2 + \frac{C_{11}h_0}{A_{11}}x +
\frac{C_{11}}{A_{11}} \right) {\cal F}_{11}(t)
\end{equation}

\noindent where

\begin{mathletters}
\label{eq:coeff_m11}
\begin{eqnarray} A_{11}&=&A_0~,~~B_{11}=B_0 \label{eq:coeff_m11_a}\\
C_{11}&=&\left[\frac{A_0h_L}{12}L^4 + \left(\frac{A_0}{3} -
\frac{h_0h_LB_0}{6}\right)L^3 -
\left(\frac{h_LB_0}{2}+h_0B_0\right)L^2 - B_0L \right]
\label{eq:coeff_m11_b} \\  {\cal F}_{11}(t)&=&\left[ \left(1-{\rm
e}^{-b^2Dt} \right) - b^2Dt\,{\rm e}^{-b^2Dt} \right]
\left[1-\Theta(t^\star)\right] \label{eq:coeff_m11_c}
\end{eqnarray}
\end{mathletters}

\noindent The contribution from that part of $R_2$ non--linear in
$m^{(0)}$, denoted previously as $m^{(12)}$, is obtained from
Eq.~(\ref{eq:m1_sol}). Making use again of the simplification
introduced earlier we obtain

\begin{eqnarray}
\label{eq:m12}
m^{(12)}(x,t)&=&-2\Gamma\left(\frac{F}{2DA_{12}}\right)^2
\left[ \frac{A_{12}^2}{30}x^6 - \frac{h_0A_{12}B_{12}}{10}x^5 +
\frac{(h_0B_{12})^2 - 2A_{12}B_{12}}{12}x^4 + \right. \nonumber \\
&& \left. \frac{h_0B_{12}^2}{3}x^3 + \frac{B_{12}^2}{2}x^2 +
\frac{h_0C_{12}}{A_{12}}x + \frac{C_{12}}{A_{12}} \right] {\cal
F}_{12}(t)
\end{eqnarray}

\noindent where

\begin{mathletters}
\label{eq:coeff_m12}
\begin{eqnarray} A_{12}&=&A_0~,~~B_{12}=B_0 \label{eq:coeff_m12_a}\\
C_{12}&=&h_L \left\{-\frac{A_{12}^2}{30} L^6 + \left(
\frac{A_{12}B_{12}h_0}{10} - \frac{A_{12}^2}{5h_L} \right) L^5 -
\frac{1}{12} \left[ (h_0B_{12})^2 - 2A_{12}B_{12} -
\frac{6A_{12}B_{12}h_0}{h_L} \right] L^4 -  \right. \nonumber \\
 & & \left. \frac{1}{3} \left[ h_0B_{12}^2 +
\frac{(h_0B_{12})^2}{h_L} - 2A_{12}B_{12} \right] L^3 - \left[
\frac{h_0B_{12}^2}{h_L} + \frac{B_{12}^2}{2} \right] L^2 -
\frac{B_{12}^2}{h_L} L \right\} \label{eq:coeff_m12_b} \\  {\cal
F}_{12}(t)&=&\left[ 1-{\rm e}^{-2b^2Dt} \right] - 2b^2Dt\,{\rm
e}^{-b^2Dt}~,~~t<t^\star \label{eq:coeff_m12_c} \\    &=&2{\rm
e}^{-b^2D(t-t^\star)} - 2b^2Dt^\star\,{\rm e}^{-b^2Dt} - \left[ {\rm
e}^{-2b^2D(t-t^\star)} + {\rm e}^{-2b^2Dt} + \right.
\nonumber \\ && \left. 2\left( {\rm e}^{-b^2Dt} - {\rm
e}^{-b^2D(2t-t^\star)} \right)
\right]~,~~t>t^\star  \label{eq:coeff_m12_d}
\end{eqnarray}
\end{mathletters}

\subsection{First Order: Dimer Density $M^{(1)}$}

In first approximation the dimer density is found from
Eq.~(\ref{eq:M1_t}); although this equation includes disassociation
effects these do not contribute directly to the solution which is
consistent with the logic underlying the approximation procedure we
are using.\cite{dimers} The initial condition for
Eq.~(\ref{eq:M1_t}) is
$M^{(1)}(x,0)\!=\!M(x,0)\!=\!0$, so that the direct contribution of
the disassociation term $\gamma M^{(1)}$ vanishes and we obtain

\begin{equation}
\label{eq:M1_sol} M^{(1)} = {\frac{1}{2}} \int_0^t {\rm d}t^\prime
{\rm e}^{-\gamma (t-t^\prime)}  \left[4m^{(0)}(x,t^\prime)/\tau +
\Gamma
\left(m^{(0)}(x,t^\prime)\right)^2\right] \end{equation}

\noindent so that $M^{(1)}$ is given directly from
Eq.~(\ref{eq:m0_simpl}) following a simple integration. If we write
$m^{(0)}\!=\!m^{(0)}(x){\cal F}_0(t)$ then, in notation identical to
that used for indicating the separate contributions to $m^{(1)}$

\begin{equation}
\label{eq:M11} M^{(11)}=2(m^{(0)}(x)/\tau) {\cal F}_{21}(t)
\end{equation}

\begin{mathletters}
\label{eq:F21}
\begin{eqnarray} {\cal F}_{21}(t)&=&\gamma^{-1}(1-{\rm e}^{-\gamma
t}) - (\gamma - b^2D)^{-1} ({\rm e}^{-b^2Dt} - {\rm e}^{-\gamma
t})~,~~t<t^\star
\label{eq:F21_a}\\   &=&\gamma^{-1} ({\rm e}^{-\gamma (t-t^\star)} -
{\rm e}^{-\gamma t}) -  (\gamma-b^2D)^{-1} ({\rm e}^{-b^2Dt} {\rm
e}^{-\gamma (t-t^\star)} -  {\rm e}^{-\gamma t})~,~~t>t^\star
\label{eq:F21_b}
\end{eqnarray}
\end{mathletters}

\begin{equation}
\label{eq:M21} M^{(12)} = \Gamma \left(m^{(0)}(x)\right)^2 {\cal
F}_{22}(t)
\end{equation}

\begin{mathletters}
\label{eq:F22}
\begin{eqnarray} {\cal F}_{22}(t)&=&\gamma^{-1}(1-{\rm e}^{-\gamma
t}) - 2(\gamma - b^2D)^{-1} ({\rm e}^{-b^2Dt} - {\rm e}^{-\gamma t})
+ \nonumber \\ && (\gamma - 2b^2D)^{-1} ({\rm e}^{-2b^2Dt} - {\rm
e}^{-\gamma t})~,~~t<t^\star \label{eq:F22_a}\\    &=&\gamma^{-1}
({\rm e}^{-\gamma (t-t^\star)} - {\rm e}^{-\gamma t}) -
2(\gamma-b^2D)^{-1} ({\rm e}^{-b^2Dt^\star} {\rm e}^{-\gamma
(t-t^\star)} -  {\rm e}^{-\gamma t}) - \nonumber\\   &&
(\gamma-2b^2D)^{-1} {\rm e}^{-\gamma t} +
(\gamma-2b^2D)^{-1}\,\left[ {\rm e}^{-2b^2D(t-t^\star)} - {\rm
e}^{-\gamma (t-t^\star)} + {\rm e}^{-2b^2Dt} \right. \nonumber\\  &&
\left. -2{\rm e}^{-b^2D(2t-t^\star)} + 2{\rm e}^{-b^2Dt^\star} {\rm
e}^{-\gamma (t-t^\star)} \right]~,~~t>t^\star \label{eq:F22_b}
\end{eqnarray}
\end{mathletters}

\subsection{Determination of the Lumped Eigenvalue $b$}
\label{subsec:eigenval}

The major approximation we have made is replacing $b_n$ by $b$ in
Eq.~(\ref{eq:m0_sol}) which is the basis for
Eqs.~(\ref{eq:m0_simpl}), (\ref{eq:m11}), and (\ref{eq:m12}). It
would be possible to use a second lumped eigenvalue in the
calculations used in obtaining the latter two results, but this
would appear to over--complicate matters at this point. What is
important for our present purposes is the prescription of a rule
that fixes $b$ in terms of the system parameters. The basis for such
a rule is a consistency requirement that the value of the coverage
as determined from Eqs.~(\ref{eq:m_t}), (\ref{eq:M_t}) and
(\ref{eq:bound}) be identical with that found directly from their
solution; we show how this can be done using the zeroth--order
approximation.

If we determine the coverage by directly integrating the solution,
$m^{(0)}$, we find

\begin{eqnarray}
\label{eq:theta}
\Theta(t)&=&\int_0^L m^{(0)}(x,t)\,{\rm d}x = \nonumber \\ &&
(FL^2/12D)
\left(4L(h_0+h_L)+h_0h_LL^2+12\right)\left(h_0+h_L+h_0h_LL\right)^{-1}{\cal
F}_0(t) \end{eqnarray}

\noindent Using Eqs.~(\ref{eq:m_t}) and (\ref{eq:bound}), after
integrating  the former we find

\begin{equation}
\label{eq:theta_t}
\frac{\partial \Theta}{\partial t} = F\!L + D\left[-h_Lm^{(0)}(L,t) -
h_0m^{(0)}(0,t) \right]
\end{equation}

\noindent from which it follows that

\begin{equation}
\label{eq:theta_sol}
\Theta(t) = (F\!L/b^2D){\cal F}_0(t)
\end{equation}

\noindent and then

\begin{equation}
\label{eq:b_sq} b^{-2} =
(L/12)\left(4L(h_0+h_L)+h_0h_LL^2+12\right)\left(h_0+h_L+h_0h_LL\right)^{-1}
\end{equation}

\subsection{Discussion}

Before considering some numerical consequences of the above
solutions, which we do in the next section, it is best to add here a
few final comments pertaining to the procedure followed in obtaining
these. The approximation procedure is {\it ad hoc\/} and follows
from physical reasoning supported by the consistency of the
resulting mathematical formulation. Although we have only considered
the first approximation beyond the step--flow solution, in principle
higher--order corrections could be obtained. It appears to be more
useful to consider instead additional effects due to moving steps
and larger islands in the present context. The former of these
appears to be tractable, and we hope to be able to report on this in
a future study. Finally, we want to emphasize that the procedure
used here is distinct from standard methods based on linearization
or a formal small parameter expansion and in our opinion captures
the essence of physical processes that are taking place better than
either of these methods.

\section{Example application: Schwoebel effect during growth and
post--growth recovery}
\label{sec:recovery}

\subsection{Surface roughness and its evolution during growth and
post--growth equilibration of the surface}

The roughness of growing surfaces has become one of the main topics
of interest in the study of surface processes. One reason for this
is the evident importance of minimizing surface roughness for
applications, e.g. for the deposition of narrow layered structures
(quantum wells, lateral superlattices, magnetic multilayers, etc.)
required for the fabrication of novel devices such as quantum well
lasers or modulation--doped field effect transistors. The
performance of such devices depends critically on the morphological
quality of epitaxial layers and interfaces.

Another reason for the intense interest in surface roughness is
purely theoretical. The roughness of growing surfaces has been
observed to exhibit asymptotic dynamical scaling behavior which has
led to the classification of growth models into various universality
classes. In a majority of the theoretical studies, the focus has
been on the evolution of the surface morphology due to fluctuations
in the incoming flux of particles and surface diffusion. However,
recently the influence of additional activation barriers to adatom
hopping between layers (the Schwoebel effect (SE),
Ref.~\onlinecite{eh,ss}) on the developing surface roughness has
been studied by Villain\cite{vill} and identified as a cause of an
intrinsic growth instability on nominally flat surfaces. On vicinal
surfaces, the additional step--edge barriers to hopping down
descending steps stabilize growth and lead to equalization of
terrace widths.\cite{ss,vill}

{}From a practical point of view, the growth interruption
technique\cite{sty} has become a useful method for obtaining
atomically abrupt interfaces during growth by MBE. In this
technique, growth is briefly interrupted by stopping the cation (but
not the anion) flux. After the growth is stopped, the surface
smoothes which causes the recovery of the specular--beam intensity
of the commonly used monitoring probe, reflection high--energy
electron--diffraction (RHEED), to its initial pre-growth value. It
has been found experimentally that the recovery of the intensity
$I(t)$ can be fitted by an empirical expression\cite{recovery}

\begin{equation} \label{eq:rec} I(t)\!=\!A_0\!+\!A_1{\rm
e}^{-t/\tau_1}\!+\!A_2{\rm e}^{-t/\tau_2}~,~~t\!>\!t^\star
\end{equation}

\noindent where $A_0$, $A_1$, and $A_2$ are constants, and $\tau_1$
and
$\tau_2$ are the time constants for the fast and slow stages of
recovery, respectively. The time constant of the initial stage of
recovery, $\tau_1$, has been shown to have a strong systematic
dependence both on the substrate temperature and on the point in
layer completion at which growth is interrupted.\cite{recovery}

The recovery process is also of profound theoretical importance. The
equilibration of a surface after a period of growth is a more
discriminating test of theoretical models than growth alone. During
growth under typical conditions, the maximum time scale is set by
the deposition flux and processes occurring over longer time scales
are ``frozen out''. However, during recovery, these processes can
come into play even though their effect during growth can be safely
omitted.\cite{vc} Indeed, Monte Carlo (MC) simulation studies which
yielded very good results when directly compared to observed
behavior during growth, appeared to be unable to reproduce the
recovery curves, in particular at low temperatures, where no
temperature dependence of $\tau_1$ was observed in
simulations.\cite{s1} Similarly, the detailed study of the recovery
process by Vvedensky and Clarke\cite{vc} using the same MC model
showed that in their simulations there was no relevant dependence of
$\tau_1$ on the point where growth was interrupted.

Peng and Whelan \cite{pw} suggested that inter-layer transport is of
key importance for the correct reproduction of the behavior of
$\tau_1$. However, large values of $\tau_1$ observed in experiment
can be explained only if there is some mechanism hindering
inter-layer transport. The obvious choice for such a mechanism is
the SE, and recent simulations\cite{sv} have shown that it indeed
results in slowing down the recovery process. However, it has been
also found that the SE alone cannot explain the observed behavior
and an additional process  must be introduced in which a freshly
arrived atom is incorporated at a site with the highest number of
nearest neighbors in a certain region around the initial site of
incidence.\cite{sv} This incorporation scheme results in smoother
edges of both pre-existing steps and adatom islands. An independent
proof exists that such a change in the model leads to a better
reproduction of the post--deposition recovery. In
Ref.~\onlinecite{mb2}, the authors had to consider detachment of
atoms from steps and break--up of islands composed of up to 10 atoms
in order to obtain good qualitative agreement with the
experimentally observed behavior. Significantly, they assumed a
shape for each island that is as close as possible to a square ({\it
i.e.\/}, very compact and with smooth edges). Similarly, the
break--up of the most stable configurations has been found to be a
crucial step for the post--growth relaxation of the surface in
Ref.~\onlinecite{kenny}.

MC simulations which included both the SE and the smoothing due to
the incorporation have led to {\it quantitative\/} agreement with the
experimental data\cite{sv} which is a strong circumstantial evidence
for the existence of the Schwoebel barriers on GaAs(001) surface.
Additional evidence has been provided in a recent paper by Johnson
{\it et al.\/}\cite{johnson} who observed (using scanning tunneling
and atomic force microscopies) growth instability and formation of
large mounds on a nominally flat GaAs(001) surface. Finally, a
recent analysis\cite{sh4} of the RHEED data of Ref.~\onlinecite{s1}
demonstrated that sticking coefficients at ascending and descending
steps on GaAs(001) surfaces are indeed strongly asymmetric providing
further support for the existence of the step--edge barriers on this
surface.

In light of the above, it would be interesting to study the influence
of the SE using the analytical results presented above and taking
advantage of the natural treatment of the boundary conditions. We
will use the parameters used for simulations of GaAs(001) growth and
try to make a comparison to some of the experimental results of
Ref.~\onlinecite{s1}.

\subsection{Adatom and dimer concentration profiles and densities,
the surface step density, and their time evolution}

In this section, we discuss adatom and dimer concentration profiles,
and the time evolution of the total densities of adatoms and dimers
and of the surface step density, during deposition and after its
cessation. We study both the case with and without the SE ({\it
i.e.\/} with different and equal sticking coefficients at boundaries
of the terraces, respectively). Note that the non--dimensionalized
form of the equations has been used in numerical evaluation of the
equations, cf. Sec.~\ref{sec:sol} and Appendix~\ref{app:nondim}.

In our calculations, we proceeded with ``deposition'' until the
saturation of the total density of adatoms and dimers on the
terraces (i.e. a steady--state situation) has been achieved, see
Fig.~\ref{fig:cov}, and then we interrupted it. The terrace width
$L$ has been chosen to be 20 lattice constants. The sticking
coefficient at the {\it ascending\/} step ($S_0$ in
Fig.~\ref{fig:model}) has been estimated based on the dimer binding
energy $E_N$ (assuming straight steps similarly to Myers--Beaghton
and Vvedensky in Ref.~\onlinecite{mb2}, Sec.~IVB), as
$S_i\!=\!1\!-\!\exp(-E_N/k_BT)$, whereas the sticking coefficient at
the {\it descending\/} step ($S_L$ in Fig.~\ref{fig:model}) was
estimated using the additional step--edge barrier height $E_B$ as
well,
$S_i\!=\!\exp(-E_B/k_BT)\!-\!\exp(-E_N/k_BT)$. The values of
parameters $E_D\!=\!1.54$~eV (which determines the diffusion
coefficient
$D$, see the paragraph following Eq.~(\ref{eq:M_t}) in
Sec.~\ref{subsec:basic_eq}), $E_N\!=\!0.231$~eV, and
$E_B\!=\!0.175$~eV were taken from the results of the fitting
procedure in Ref.~\onlinecite{sv}. The attempt frequency $\nu_0$ has
been calculated from $\nu_0\!=\!2k_BT/h$, where $h$ is the Planck's
constant. The value of the flux $F$ was 0.20 ML/s, and the values of
the substrate temperature $T$ are given in the figures.

Adatom and dimer concentration profiles shown in
Fig.~\ref{fig:profiles} have been calculated at the time $t^\star$
when the deposition was stopped (in the saturation region). The
upper two panels ({\bf a} and {\bf b}) of  Fig.~\ref{fig:profiles}
are for the case without the SE, whereas the lower two ({\bf c} and
{\bf d}) for the case with the SE. The panels {\bf a} and {\bf c}
show both the total adatom density $m$ (solid lines) and the adatom
density in the lowest order of approximation, $m^{(0)}$ (dotted
lines), cf. Eqs.~(\ref{eq:m0_simpl}) and ~(\ref{eq:coeff_m0}).

The profiles for the case with the SE are asymmetric and exhibit a
maximum near the reflecting boundary. Notice that significantly
higher adatom and dimer densities build up on the terraces due to
the presence of the Schwoebel barriers. It is also interesting to
mention that whereas the densities with and without the SE are of
approximately the same order of magnitude just after the growth is
stopped (Fig.~\ref{fig:profiles}) they differ by many orders of
magnitude after the equilibration due to a  larger time constant for
this process in the case with the SE (see below). Note also that
concentration profiles do not vanish at boundaries,
$x\!=\!0,1$ (in contrast to those in Ref.~\onlinecite{mb1,lzvw}) due
to the microscopic treatment of the boundary conditions in our
approach, cf. Sec.~\ref{subsec:basic_eq}.

In Fig.~\ref{fig:cov}, the time evolution of the total densities of
adatoms and dimers on terraces is shown. The strong temperature
dependence of the densities and of the difference between $m$ and
$m^{(0)}$ can be again observed. However, the most significant
difference is a rather drastic change of the time for the decay of
the adatom and dimer densities due to the SE. This confirms that the
inclusion of the barriers to hopping at step edges does lead to a
slower relaxation of the surface after growth is terminated, a
conclusion reached previously in the MC study mentioned
above.\cite{sv}

Finally, in Fig.~\ref{fig:ssd}, the evolution of the surface step
density calculated in a pseudo--two--dimensional approximation
(similar to that used in Ref.~\onlinecite{mb2}) is shown. For the
calculation of this quantity, we supposed that every free adatom and
dimer contributes 4 and 6 steps, respectively. The panel {\bf a}
shows the results without the SE, whereas the panel {\bf c} those
with the SE. The experimental results at similar temperatures taken
from Ref.~\onlinecite{s1} (RHEED specular--beam intensity during the
post--growth recovery on a GaAs(001) vicinal surface misoriented by
$2^\circ$ toward {[}010{]} direction) are shown in the panel {\bf
b}. Notice the order of magnitude difference in the time scale which
is discussed in the following section.

\subsection{Discussion}

The above numerical results are based on evaluation of the preceding
analytical results and agree with earlier numerical
studies.\cite{mb1,mb2,fuen,lzvw} The advantage of our approach
(besides the important fact we were able to obtain analytic
formulae) is the microscopic level of the description of the growth
process and resulting natural treatment of  boundary conditions. We
have demonstrated than even a relatively small Schwoebel barrier has
a profound influence on behavior during growth and, in particular,
during the post--growth recovery. Another detail worth mentioning is
that had we used
$\tau_0$ instead $\tau_r$ for the time constant in the prefactor
$\Gamma$ in Eq.~(\ref{eq:M_t}) (see the discussion in
Sec.~\ref{subsec:basic_eq}), the contribution
$m^{(12)}$ to the adatom density would be underestimated by several
orders of magnitude. Let us now discuss several problems related to
our results.

As expected, the difference between the adatom concentration profiles
obtained using formulae for $m$ and $m^{(0)}$
(Fig.~\ref{fig:profiles} {\bf a}, {\bf c}) is not important under
conditions near the step--flow growth mode (the biggest difference
being observed near the maxima), but increases as the temperature is
decreased (cf. similar results in Fig.~1 of Ref.~\onlinecite{lzvw}).
However, this does not mean that the dimer (and larger islands)
formation is not important. It was shown by Stoyanov\cite{stoyanov}
and Myers--Beaghton and Vvedensky\cite{mb1} that adatom interactions
lower the adatom diffusivity by several orders of magnitude and, if
not taken into account properly (as it is the case when the Einstein
relation for surface diffusion
$x^2\!=\!2Dt$ is used\cite{n}), lead to an estimate of the critical
temperature $T_c$ (at which growth starts to proceed via step flow)
which is incorrect by up to two hundred kelvin.\cite{mb1} In other
words, whereas the concentration profiles can be calculated in the
lowest order of approximation near $T_c$, the critical temperature
itself must be determined taking higher--order effects into account.

The time constants for the post--growth recovery of the adatom and
dimer densities are an order of magnitude smaller than observed in
experiment (cf. Fig.~\ref{fig:cov} and particularly
Fig.~\ref{fig:ssd}). This is not surprising given the fact that in
our model, the effect of break--up of larger islands is not taken
into account. It was convincingly demonstrated by Myers--Beaghton
and Vvedensky\cite{mb2} and Kenny {\it et al.\/}\cite{kenny} how
significant these processes are for the correct reproduction of the
recovery curves, and this process was also shown to be important in
the MC study mentioned above\cite{sv} where it was found that step
and island edges smoothed by the incorporation mechanism of the
incoming atoms (see above) act in concert with the step--edge
barriers to slow down the recovery process. Also the fact that
one--dimensional diffusion in our model leads to more rapid changes
of the surface adatom population as compared to the two--dimensional
case adds to the difference in the speed of surface smoothing.

Finally, in the framework of our simple model, we cannot observe
oscillations of the density of adatoms, dimers, or the surface step
density (or oscillations of related quantities like the
specular--beam intensity of some diffraction probe calculated in the
kinematic approximation). Such oscillations are not experimentally
observed near the step--flow growth mode where our model is,
strictly speaking, only supposed to be valid. However, it is
possible to obtain oscillations at lower temperatures (or higher
fluxes) even in this class of models provided the step movement is
taken into account.\cite{mb2} As we already mentioned, the inclusion
of this effect into our perturbative scheme seems to be possible
(cf. Sec.~\ref{sec:intro}).

\section{Summary}
\label{sec:summary}

In summary, we have investigated epitaxial growth on vicinal
(stepped) surfaces using an analytical program based on a
perturbative solution of the generalized Burton--Cabrera--Frank
equations. The non--standard scheme followed by us is motivated by
physical insight into the growth process. This approach allowed us
to obtain closed--form analytical solutions for the adatom and dimer
densities while using a microscopic description of the boundary
conditions. The scheme is open to further improvements and inclusion
of other effects such as the movement of the pre-existing steps.

We applied the analytical results to the study of the influence of
the Schwoebel effect (additional barriers to inter--layer hopping at
step edges) on the evolution of adatom and dimer populations and of
the surface step density during growth and post--growth
equilibration of the surface and compared them (where it was
possible) to the experimental results. The Schwoebel effect was
shown to have a significant impact on the behavior during growth
and, in particular, during the post--growth recovery, but it has
also become apparent that larger islands and their decay have to be
taken into account if the recovery curves obtained experimentally
and in simulations are to be reproduced. In order to obtain a more
realistic description of the growth process at lower temperatures
(including  oscillations of step density or kinematic intensity),
additional processes must be considered such as the above--mentioned
step movement.

\acknowledgments

The NATO Travel Grant CRG 931508, and the support of Imperial
College and the Research Development Corporation of Japan under the
auspices of the ``Atomic Arrangement:~Design and Control for New
Materials'' Joint Research Program are gratefully acknowledged. S.H.
also thanks the Department of Civil Engineering and Operations
Research at Princeton University for their hospitality.

\appendix
\section{Dimensionless variables}
\label{app:nondim}

It is convenient for the purposes of assessing the effects of various
parameters to represent the results obtained in dimensionless form.
Introducing $x^\prime\!=\!xL$ and $({h_i})^\prime\!=\!ah_i$ and
using the relationships $D\!=\!a^2/2\tau_0$,
$a/\tau_0\!=\!\sqrt{2\alpha}$ which provide reasonable estimates, we
then find, e.g. for $m^{(0)}$

\begin{equation} a^2m^{(0)} = (\tau_0/\tau A_0) [ -A_0(L/a)^2x^2 +
h_0B_0(L/a)x+B_0]{\cal F}_0(t)
\end{equation}

\noindent where we have dropped the primes for notational
convenience. Now
$0\!\le\!x\!\le\!1$ and

\begin{eqnarray} h_i&=&\left[S_i/(2-S_i)\right](2/\sqrt{\pi}) \\
A_0&=&h_0+h_1+h_0h_1(L/a) \\ B_0&=&h_1(L/a)^2+2(L/a)
\end{eqnarray}

\noindent and

\begin{equation} 1/Db^2 =
(L/a)\,(\tau_0/6A_0)\,[4(L/a)(h_0+h_1)+h_0h_1(L/a)^2+12]
\end{equation}

\noindent Similar results follow for $m^{(1)}$ and $M^{(1)}$ but
because they are so lengthy we do not include them here.

\begin{figure}
\caption{Schematic picture of the processes considered in our model.
Atoms are deposited (1) at a rate $F$ onto the substrate with a
lattice constant $a$ held at a temperature $T$. They migrate (2)
with the diffusion coefficient $D$ and attach either to the
pre-existing steps (with probabilities given by the sticking
coefficients $S_0$ and $S_L$) or form dimers (3), (4), which can
subsequently disassociate (5). Two ways of a dimer creation, by a
deposition of an incoming atom onto a nearest--neighbor site of a
surface adatom (3) or by an encounter of two migrating adatoms (4),
are shown.}
\label{fig:model}
\end{figure}

\begin{figure}
\caption{Adatom ({\bf a}, {\bf c}) and dimer ({\bf b}, {\bf d})
concentration profiles immediately after growth is stopped at the
steady--state region. The panels {\bf a}, {\bf b} and {\bf c}, {\bf
d} show the results obtained without and with the Schwoebel effect,
respectively. See text for more details.}
\label{fig:profiles}
\end{figure}

\begin{figure}
\caption{Time evolution of adatom ({\bf a}, {\bf c}) and dimer ({\bf
b}, {\bf d}) densities. The panels {\bf a}, {\bf b} and {\bf c},
{\bf d} show the results obtained without and with the Schwoebel
effect, respectively.}
\label{fig:cov}
\end{figure}

\begin{figure}
\caption{Time evolution of the surface step density (panels {\bf a},
{\bf c}) and the RHEED specular beam intensity ({\bf b}). The panels
{\bf a} and {\bf c} show the results obtained without and with the
Schwoebel effect, respectively. The direction of vertical axis in
the plots of the step density is reversed to allow for comparison
with the RHEED specular--beam intensity (see
Refs.~\protect\onlinecite{s1,sv}), {\it i.e.\/} the step density
increases {\it downwards\/}.}
\label{fig:ssd}
\end{figure}


\begin{references}
\bibitem{fuen} V. Fuenzalida,  Phys. Rev. B {\bf 44}, 10835 (1991).
Only the steady state is treated here but the generalization to the
time--dependent case is straightforward.
\bibitem{step-flow}  It is useful to note that the term ``step flow''
is usually used for a situation when no oscillations of a
surface--sensitive probe (e.g. RHEED) are observed.  An example of
such experimental conditions taken from Ref.~\onlinecite{s1} for
growth on a vicinal GaAs(001) surface misoriented by $2^\circ$
towards {[}010{]} direction is $T\!=\!850$~K at $F\!=\!0.20$~ML/s.
However, large islands are still present on the surface under such
conditions, but they are incorporated into advancing steps before
they can nucleate a new layer. Therefore, a higher temperature or a
lower incoming flux would be necessary to achieve ``ideal''
conditions for our model.
\bibitem{ori} This is the case for a (001) surface. Other choices are
possible in which case the coefficients 8 ($=\!2\times4$) and 4 in
Eqs.~(\ref{eq:m_t}) and (\ref{eq:M_t}) must be changed accordingly.
\bibitem{dimers} By this we mean that these effects are not included
in the $m^{(0)}$ term indicated in $R$ in Eq.~(\ref{eq:M1_t}).
However, by retaining the $M^{(1)}$ in that term we are able, in
part, to include the effects of disassociation in a self--consistent
manner without having to carry out a higher--order (and difficult)
calculations.
\end{references}
\end{document}